\documentclass[useAMS,usenatbib]{mn2e}
\usepackage{graphicx,amssym}
\newcommand{\bc}{\begin{center}}
\newcommand{\ec}{\end{center}}

\title[How do galaxies populate Dark Matter halos?]
      {How do galaxies populate Dark Matter halos?}
\author[Qi Guo, Simon White, Cheng Li, Michael Boylan-Kolchin]
       {Qi Guo$^{1}$ \thanks{Email: guoqi@mpa-garching.mpg.de}, Simon White$^1$, Cheng Li$^{1,2}$, Michael Boylan-Kolchin$^{1}$
        \\     
	\\
        $^{1}$ Max Planck Institut f\"ur
         Astrophysik, Karl-Schwarzschild-Str. 1, 85741 Garching, Germany\\
        $^{2}$ MPA/SHAO Joint Center for Astrophysical Cosmology at Shanghai Astronomical Observatory, 80 Nandan Road, Shanghai 200030, China }
\begin{document}

\date{Accepted  ???? ??. 2009 ???? ??}

\pagerange{\pageref{firstpage}--\pageref{lastpage}} 
\pubyear{200?}

\maketitle

\label{firstpage}

\begin{abstract}
For any assumed standard stellar Initial Mass Function, the Sloan
Digital Sky Survey (SDSS) gives a precise determination of the
abundance of galaxies as a function of their stellar mass over the
full stellar mass range $10^8 M_{\odot} < M_* < 10^{12}
M_{\odot}$. Within the concordance $\Lambda$CDM cosmology, the
Millennium simulations give precise halo abundances as a function of
mass and redshift for all halos within which galaxies can form.  Under
the plausible hypothesis that the stellar mass of a galaxy is an
increasing function of the maximum mass ever attained by its halo,
these results combine to give halo mass as a function of stellar
mass. The result agrees quite well with observational estimates of
mean halo mass as a function of stellar mass from stacking analyses of
the gravitational lensing signal and the satellite dynamics of SDSS
galaxies. For $M_* \sim 5.5\times 10^{10} M_{\odot}$, the stellar mass
usually assumed for the Milky Way, the implied halo mass is $\sim
2\times 10^{12} M_{\odot}$, consistent with most recent
direct estimates and inferences from the MW/M31 Timing Argument.  The fraction of the baryons associated with each
halo which are present as stars in its central galaxy reaches a
maximum of 20\% at masses somewhat below that of the Milky Way, and
falls rapidly at both higher and lower masses. These conversion
efficiencies are lower than in almost all recent high-resolution
simulations of galaxy formation, showing that these are not yet viable
models for the formation of typical members of the galaxy
population. When inserted in the Millennium-II Simulation, our derived
relation between stellar mass and halo mass predicts a stellar mass
autocorrelation function in excellent agreement with that measured
directly in the SDSS. The implied Tully-Fisher relation also appears
consistent with observation, suggesting that galaxy luminosity
functions and Tully-Fisher relations can be reproduced simultaneously
in a $\Lambda$CDM cosmology.
\end{abstract}


\section{Introduction}
\label{sec:gfintro}
It has long been known that baryons do not dominate the mass in the
Universe, and recent observations of microwave background fluctuations
have demonstrated that only about 15\% of cosmic matter is in the form
of baryons. The remaining 85\% is apparently made up of some as yet
unidentified, weakly interacting, non-baryonic particle, so-called
cold dark matter \citep{Komatsu2009}. Together these two matter
components account for only a quarter of the current energy density of
the Universe; the rest comes from a near-uniform dark energy field
which drives the observed acceleration of the cosmic
expansion \citep{Astier2006}. Nonlinear structure formation in this
concordance $\Lambda$CDM cosmology proceeds through gravitationally
driven hierarchical collapse and aggregation. Galaxies form by the
cooling and condensation of gas at the centres of an evolving
population of dark matter halos, as originally set out in a different
context by \cite{White1978}.

The abundance of dark matter halos as a function of their mass can be
predicted by the simple theory of \cite{Press1974} and its extensions,
and is extremely broad. For example, halos with mass exceeding
$10^{14}M_\odot$ are predicted to contain roughly 10\% of all dark
matter in today's Universe; halos with mass below $10^{-5}M_\odot$
should contain another 10\%; and the remaining 80\% is distributed
over the intervening 19 orders of magnitude, with the median point near
$10^{10}M_\odot$ \citep[e.g.][]{Mo2002,Angulo2009}. This behaviour has
been confirmed by detailed numerical simulations of cosmic evolution
over the limited mass range accessible to them (currently
$>10^{9}M_\odot$; see below).  The observed distribution of stars among
galaxies is much more confined, however, with 10\% of stars in
galaxies with stellar mass above $1.3\times 10^{11}M_\odot$, 10\% in
galaxies with stellar mass below $5\times 10^{9}M_\odot$ and a median
point near $4\times 10^{10}M_\odot$ \citep{Li2009}. This implies that
the baryons are converted into stars with very different efficiencies
in halos of different mass. Star formation is most efficient at the
centres of halos of typical galaxies such as the Milky Way, and is
substantially less efficient at the centres of much more massive or
much less massive halos \citep{Navarro2000}. \cite{Larson1974} noted
that star formation would likely be inefficient in objects with
low escape velocities because of the ease with which winds can expel gas, and \cite{White1978} invoked this process to explain the
relatively small fraction of all stars which end up in low-mass galaxies.
The inefficiency of star formation at the centre of massive halos is
related to the well known ``cooling flow paradox''
\citep[e.g.][]{Fabian2001}. Recent work suggests that it may result
from feedback from the central radio AGN \citep{Tabor1993,Ciotti2001,
Birzan2004,Croton2006,Bower2006}.

There are several ways to study how galaxies populate dark matter
halos. The most straightforward is direct simulation of galaxy
formation in its cosmological context. An $N$-body treatment of the
evolution of the dark matter component is combined with either a
hydrodynamical \citep[e.g.][]{Cen2000, Springel2003, Keres2005,
Pfrommer2006, Sijacki2007} or a semi-analytic \citep[e.g.][]
{Kauffmann1999, Springel2001, Hatton2003, Springel2005b, Kang2005}
treatment of baryonic evolution. The advantage of these methods is
that they track galaxy and dark matter halo evolution across cosmic
time in a physically consistent way, providing positions, velocities,
star formation histories and other physical properties for the galaxy
populations of interest.  In recent years both techniques have had
considerable success in reproducing observations. Hydrodynamical
simulations provide a much better description of diffuse gas
processes, but are relatively inflexible, typically producing galaxy
populations which are at best a rough match to observation.  The
greater flexibility and speed of semi-analytic methods allows much
better reproduction of observational data at the expense of a
schematic treatment of diffuse gas physics.  Both schemes implement
simplified and highly uncertain recipes to treat star and black hole
formation and related feedback processes. For both, these entrain
considerable uncertainty about whether the physics of galaxy formation
is reliably represented.

Even better fits to the observed luminosity, colour and clustering
distributions of galaxies can be obtained by giving up on any attempt
to represent formation physics, and instead using simple statistical
models with adjustable parameters to populate dark matter halos with
galaxies. By adjusting model assumptions and their associated
parameters, this halo occupation distribution (HOD) approach is able
to match observed statistics like galaxy luminosity and correlation
functions as a function of luminosity and colour with remarkable
accuracy within the concordance $\Lambda$CDM cosmology
\citep[e.g.][]{Cooray2002,Berlind2002,Yang2003}. Their disadvantage is
that they use no information about the evolution of a system when
populating it with galaxies, and this formation history may have
significant influence on the properties of the
galaxies \citep[e.g.][]{Gao2005,Croton2007}. Several authors have
explored schemes intermediate between direct simulation and HOD
modelling in an attempt to retain some of the advantages of each
\citep[e.g.][]{Wang2006, Wang2007, Conroy2009}.

An alternative is to link galaxies to their halos/subhalos by matching
observed galaxy luminosity functions to simulated halo mass functions
assuming a unique and monotonic relation between galaxy luminosity and
halo mass. This method was proposed by \cite{Vale2004} and then
extended by several groups to consider a variety of properties both
for the galaxies and for their halos/subhalos
\citep{Conroy2006, Shankar2006, Conroy2007, Baldry2008, Moster2009}.
In particular, \cite{Moster2009} show how the method can be extended
to allow a scatter in the properties of galaxies associated with halos
or subhalos of given mass \citep[see also][]{Wetzel2009}.

Here we follow this latter approach. We combine a precise stellar mass
function based on the full spectroscopic dataset of the most recent
SDSS data release \citep{Li2009} with a precise halo/subhalo mass
function obtained from the Millennium Simulation
(MS; \citealt{Springel2005b}) and the higher resolution Millennium-II
Simulation (MS-II; \citealt{Boylan2009}).  This yields a much more
accurate relation between galaxy stellar mass and dark matter halo
mass than could be derived from earlier data. We compare this relation
with direct observational estimates of the mean mass of halos
surrounding galaxies of given stellar mass inferred from gravitational
lensing and satellite galaxy dynamics data. We also compare the halo
masses predicted for the Milky Way and other Local Group galaxies with
estimates derived from dynamical data. As a further consistency test
we populate halos/subhalos in the MS and MS-II with galaxies of
stellar mass chosen according to our relation, and we compare the
stellar mass correlation function of the result with the SDSS
measurement of \cite{Li2009}. We derive the star formation
efficiencies implied as a function of halo mass by our relation, and
we compare them to the efficiencies in published hydrodynamical
simulations of galaxy formation in the $\Lambda$CDM cosmology.
Finally, we revisit the issue of whether the abundances and circular
velocities of galaxies can be fit simultaneously in a $\Lambda$CDM 
cosmology \citep[e.g.][]{Cole2000}.
 
In the next section we briefly describe the two Millennium
simulations, and the definitions of halo mass that we adopt for the
rest of this paper. The relation between galaxy stellar mass and dark
matter halo mass is derived at the beginning of our results section,
Sec.~\ref{sec:resultsgf}, and in later subsections we compare it with
direct observational determinations, we give predictions for the halo
masses of Local Group galaxies, we compare the implied stellar mass
correlation function with that measured in SDSS, and we revisit the
problem of simultaneously reproducing the luminosity function and the
Tully-Fisher relation in a hierarchical cosmology. Conclusions and a
discussion of our results are presented in Sec.~\ref{sec:conclusiongf}.

\section{Dark Matter Halos}
\label{sec:simu}
Both the \emph {Millennium Simulation} and the \emph {Millennium-II
Simulation} adopt the concordance $\Lambda$CDM cosmology with
parameters chosen to agree with a combined analysis of the
2dFGRS \citep{Colless2001} and the first-year WMAP data
\citep{Spergel2003}. The parameters are $\Omega_{\rm
m}=0.25$, $\Omega_{\rm b}=0.045$, $H_0=73 {\rm km s}^{-1}{\rm
Mpc}^{-1}$, $\Omega_\Lambda=0.75$, $n=1$, and $\sigma_8=0.9$.  These
parameters are only marginally consistent with analysis of the latest
WMAP and large-scale structure data \citep{Komatsu2009} but the
differences are too small to significantly affect the analysis of this
paper. Both simulations use 2160$^3$ dark matter particles to trace
evolution from z $\sim$ 127 to z $\sim$ 0. They were carried out in
cubic volumes with periodic boundary conditions and sides of length
685 Mpc and 137 Mpc for the MS-I and MS-II, respectively,
corresponding to particle masses of $1.2\times 10^{9}M_{\odot}$ and
$9.5\times 10^{6}M_{\odot}$. The large volume of the MS enables one to
study even the cD galaxies of rare and massive clusters with good
statistical power, while the excellent mass resolution of the MS-II
can resolve the dark matter halos predicted to host even the faintest
known dwarf galaxies.

For each of the output dumps, friends-of-friends (FOF) groups are
identified in each simulation by linking together particles separated
by less than 0.2 of the mean interparticle
separation \citep{Davis1985}. The SUBFIND
algorithm \citep{Springel2001} was then applied to each FOF group in
order to split it into a set of disjoint, self-bound subhalos, which
represent locally overdense and dynamically stable subunits within the
larger system. The main subhalo is defined as the most massive
self-bound subunit of a FOF group and normally contains most of its
mass. Merger trees were then built which link each subhalo present in
a given dump to a unique descendent in the following dump. These allow
us to track the formation history of every halo/subhalo present at
$z=0$. We refer readers to \cite{Springel2005b} and \cite{Boylan2009}
for a more detailed descriptions of these simulations and
post-processing procedures.
 
\begin{figure}
\bc
\hspace{-0.6cm}
\resizebox{8.5cm}{!}{\includegraphics{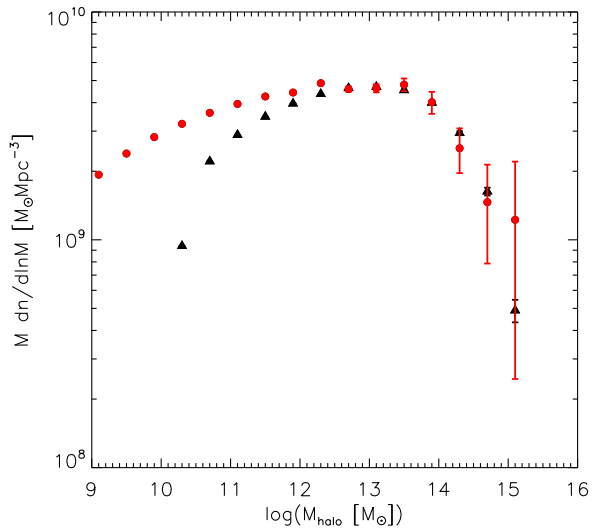}}\\%
\caption{Dark matter halo mass functions at $z=0$ where ``halo'' is
defined to include both main subhalos and satellite subhalos. Halo
mass, $M_{halo}$, is defined as the current virial mass for main
subhalos and as the virial mass immediately prior to accretion for
satellite subhalos. In both cases this is normally the maximum mass
attained over the subhalo's history.  Black triangles are for the MS and
red dots are for the MS-II. Poisson error bars based on halo counts
are shown for both simulations. }

\label{fig:MFwSat}
\ec
\end{figure}

In this work, we assume that both main subhalos and satellite subhalos
have galaxies at their centres, and that the stellar masses of these
galaxies are directly related to the {\it maximum} dark matter mass
ever attained by the subhalo during its evolution. We denote this mass
by $M_{halo}$.  In practice this mass is usually the mass at $z=0$ for
main subhalos, and the mass just prior to accretion for satellite
subhalos. Semi-analytic simulations show that for satellite systems
this latter mass is much more closely related to the stellar mass of
the central galaxy than is the $z=0$ mass of the subhalo, because the
latter has often been very substantially reduced by tidal
stripping \citep{Gao2004, Wang2006,
Font2008}. \cite{Vale2004}, \cite{Conroy2006} and \cite{Berrier2006}
present general plausibility arguments for such an assumption rather
than studying any specific galaxy formation model. We thus need to
estimate the abundance of (sub)halos in the Millennium simulations as
a function of this $M_{halo}$.

For each FOF group, we define the centre as the minimum of the
gravitational potential well and we define the virial radius,
$R_{vir}$, as the radius that encloses a mean overdensity of 200 times
the critical value. The mass within $R_{vir}$ is then defined as the
virial mass:
\begin{equation}
M_{halo}=\frac{100}{G}H^2(z)R^3_{vir}.
\end{equation}
We define $M_{halo}$ for a main subhalo to be its current virial mass,
and for a satellite subhalo to be its virial mass immediately prior to
accretion onto a larger system, i.e. its virial mass immediately
before it last switches from being a main subhalo to a satellite
subhalo. Hereafter, we refer to both main subhalos and satellite
subhalos as ``halos'' and we refer to $M_{halo}$ defined in this way
as the ``halo mass''.

Fig.\ref{fig:MFwSat} shows halo mass functions for the two Millennium
simulations at $z= 0$. Black triangles refer to the MS and red dots to the
MS-II. The two simulations agree well above $10^{12.3}M_{\odot}$ but
below this threshold, the MS lies progressively below the MS-II. This
is due to resolution effects, which set in at substantially higher
masses than for the FOF halo mass function in Fig.~9
of \cite{Boylan2009} because of the inclusion of satellite subhalos.
These can fall below the resolution of the MS at $z=0$ yet still be
relatively massive at the time of infall.  In the following, we
combine the part of the MS mass function with $M_{halo}>1.9\times
10^{12}M_{\odot}$ with the part from the MS-II with $M_{halo}< 1.9\times
10^{12}M_{\odot}$ in order to represent the overall dark halo mass
function as well as possible. Based on the deviations between the two
simulations visible in Fig.\ref{fig:MFwSat}, we estimate that the
resulting function should be accurate to better than about 10\% from
$10^{10}M_\odot$ up to $10^{15}M_\odot$. This will turn out to cover
the full halo mass range of relevance for real galaxies.
\section{Galaxy Formation Efficiency}
\label{sec:resultsgf}
\subsection{Connecting Galaxies to Dark Matter Halos}
We connect dark halo mass $M_{halo}$ to the stellar mass of the
central galaxy by assuming a one-to-one and monotonic relationship
between the two. In practice, if the number density
of dark matter halos with mass exceeding $M_{halo}$ matches the number density
of galaxies with stellar mass exceeding $M_{*}$,
\begin{equation}
n(>M_{halo}) = n(>M_{*}),
\label{eq:matching}
\end{equation}
then we assume galaxies of stellar mass $M_{*}$ to reside at the
centre of dark matter (sub)halos of mass $M_{halo}$.
\begin{figure}
\bc
\hspace{-0.6cm}
\resizebox{8.5cm}{!}{\includegraphics{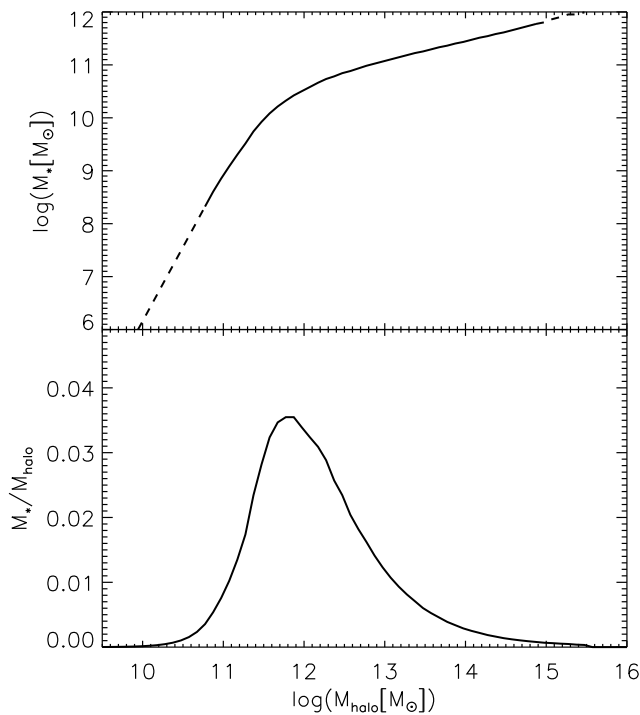}}\\%
\caption{The stellar mass -- dark matter halo mass relation. The solid
curve is obtained by matching galaxy abundances from SDSS/DR7 to dark
matter halo abundances from the combination of the MS and the MS-II
(Fig.\ref{fig:MFwSat}). The dashed curve shows an extrapolation of
this relation to stellar masses of $10^6M_{\odot}$ and
$10^{12}M_{\odot} $ at the low- and high-mass ends, respectively. The
bottom panel shows the ratio of stellar mass to halo mass as a
function of halo mass.}
\label{fig:mstar2mhalo}
\ec
\end{figure}

To derive the relation between $M_{halo}$ and $M_*$ we need to combine
the halo mass function of Fig.~\ref{fig:MFwSat} with an equally
precise observed stellar mass function for galaxies.  We take the
recent measurement presented by \cite{Li2009}. This is based on a
complete and uniform sample of almost half a million galaxies from the
Sloan Digital Sky Survey data release 7 (SDSS/DR7)
\citep{York2000,Abazajian2008}. This extends over almost four
orders of magnitude in stellar mass ($10^8 M_{\odot} - 10^{11.7}
M_{\odot}$) with very small statistical errors. The main residual
uncertainty comes from possible systematic errors in the determination
of stellar masses from the SDSS photometry. Here we convert from
masses based on SDSS $r$-band Petrosian luminosities, as used
by \cite{Li2009}, to masses based on SDSS $r$-band ``model''
luminosities. The latter are generally thought to give a better
estimate of the {\it total} luminosity of galaxies. This conversion is
discussed in detail in Appendix A, which also gives a modified version
of the fitting formula of \cite{Li2009} which represents this ``total
stellar mass'' function. The correction increases stellar masses by
about 9\% on average. If we leave aside uncertainties in the stellar
Initial Mass Function then results in Appendix A and in the appendices
of \cite{Li2009} suggest that the remaining systematic uncertainty in
the stellar mass functions are of order 10\% in stellar mass. Purely
statistical errors are much smaller than this. Since the abundances
matched in Eq.~\ref{eq:matching} range over almost six orders of
magnitude, such uncertainties have only a very small effect on the
$M_{halo}$ -- $M_*$ relation.

Our relation between galaxy stellar mass and the dark matter halo mass
is shown in the upper panel of Fig.~\ref{fig:mstar2mhalo}. The solid
curve uses SDSS/DR7 data over the stellar mass range from $10^{8.3}$
to $10^{11.8}M_{\odot}$, which corresponds to dark matter halo masses
between $10^{10.8}M_\odot$ and $10^{14.9}M_{\odot}$. We extrapolate
this relation to $10^6M_{\odot}$ at the low-mass end and to
$10^{12}M_{\odot} $ at the high-mass end assuming constant slope, as
indicated by the dashed extensions. Galaxies with mass around
$10^6M_{\odot}$ are expected to reside in dark matter halos with mass
$\sim 10^{10}M_{\odot}$, where we expect errors in our abundance
estimates still to be below 10\%.  At the high-mass end, the stellar
mass of the central galaxy becomes very insensitive to its dark matter
halo mass, indicating a suppression of star formation in the cores of
halos more massive than $\sim 10^{13}M_{\odot}$.

If we adopt the functional form suggested by \cite{Yang2003}
and \cite{Moster2009}, our derived relation can be approximated to high
accuracy by
\begin{equation}
\frac{M_*}{M_{halo}}=c\times \left[ \left( \frac{M_{halo}}{M_0}\right) ^{-\alpha} +\left(\frac{M_{halo}}{M_0}\right)^{\beta}\right]^{-\gamma}
\end{equation}
where $c=0.129$, $M_0=10^{11.4}M_{\odot}$, $\alpha=0.926$,
$\beta=0.261$ and $\gamma=2.440$. Note that this formula has been
fitted to our results over the halo mass range $10^{10.8}$ to
$10^{14.9}M_{\odot}$, corresponding to the solid curve in the upper
panel of Fig.~\ref{fig:mstar2mhalo}.

In the bottom panel of Fig.~\ref{fig:mstar2mhalo} we show the ratio of
stellar mass to dark halo mass as a function of halo mass.  This
reaches a maximum in halos with $M_{halo}\sim 10^{11.8}M_{\odot}$,
slightly less massive than the halos which host $L^*$ galaxies. The
peak value is around 3.5\%. The ratio drops very rapidly towards both
lower and higher halo masses: $M_*/M_{halo} < $ 0.27\% in dark matter
halos with mass $~\sim 10^{10.7}M_{\odot}$ and $M_*/M_{halo} \sim $
0.09\% in clusters with $~\sim 10^{14.8}M_{\odot}$. (Note that in the
latter case the stellar mass refers {\it only} to the central galaxy.)

Semi-analytic models like that of De Lucia \& Blaizot (2007; DLB07)
produce curves very similar to those of Fig. 2 but noticeably
offset. This offset comes from several sources. As may be seen in
Fig.1, the MS does not produce the correct (sub)halo abundance below
$M_{halo} = 10^{12}M_{\odot}$, so that a SAM based on the MS alone (like that of
DLB07)is skewed as a result. At the moment, there are no semi-analytic
models tuned to work simultaneously on the MS and MS-II simulations,
though we intend to produce such models in the future (Guo et al., in
preparation). In addition, the DLB07 models do not accurately fit the
\cite{Li2009} mass function (see their Fig. B1) so this also
introduces an appreciable offset in the $M_*$ - $M_{halo}$ relation. The scatter
of this model around its own $M_*$ - $M_{halo}$ relation is, however, quite
small and is comparable with the values we test below.

\subsection{Comparison to observed $M_{halo}$ - $M_{*}$ relation}
\begin{figure}
\bc
\hspace{-0.6cm}
\resizebox{8.5cm}{!}{\includegraphics{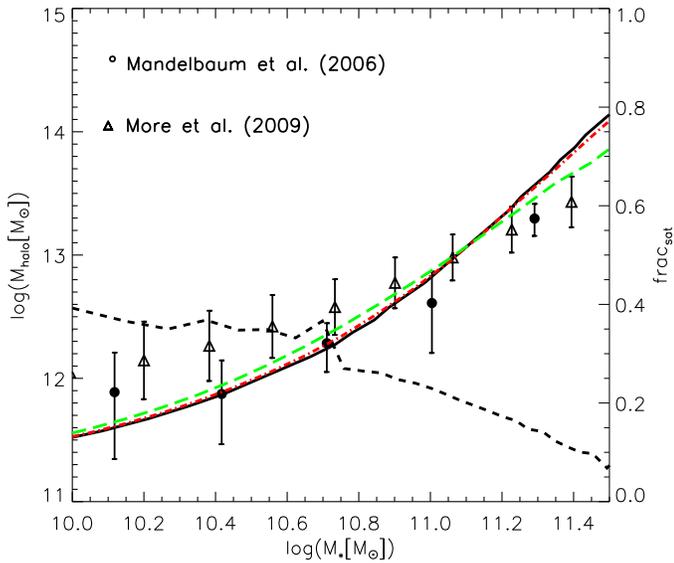}}\\%
\caption{Dark matter halo mass as a function of stellar mass. The
thick black curve is the prediction from abundance matching assuming no
dispersion in the relation between the two masses. Circles with error
bars are weak lensing estimates of the mean halo mass of central
galaxies as a function of their stellar mass \citep{Mandelbaum2006}.
The error bars show the 95\% confidence ranges. Triangles with
$1\sigma$ error bars show mean halo masses as a function of central
galaxy stellar mass derived from the stacked kinematics of satellite
galaxies \citep{More2009}. Red and green dashed curves show
abundance matching predictions for mean halo mass as a function of
galaxy stellar mass assuming dispersions of 0.1 and 0.2, respectively,
in $\log M_*$ at given halo mass. The dashed black curve is the
satellite fraction as a function of stellar mass, as labelled on the
axis at the right-hand side of the plot.}
\label{fig:mstar2mhaloobs}
\ec
\end{figure} 

We now focus on galaxies with stellar mass between $10^{10}M_{\odot}$
and $3\times 10^{11}M_{\odot}$ and show halo mass vs. stellar mass in
Fig.~\ref{fig:mstar2mhaloobs}. The solid curve is the prediction of
our abundance matching as shown already in Fig.~\ref{fig:mstar2mhalo}.
The circles with error bars show mean halo mass as a function of
central galaxy stellar mass as obtained from gravitational lensing
measurements \citep{Mandelbaum2006}. The data consisted of 351,507
galaxies from SDSS, including both early- and late-type galaxies in
the mass range [$0.7\times 10^{10}M_{\odot}$, $4\times
10^{11}M_{\odot}$]. The error bars indicate 95\% confidence
intervals. The stellar masses used here are based on the photometric
properties of SDSS galaxies (see Appendix A), while those
in \cite{Mandelbaum2006} were estimated by \cite{Kauffmann2003} based
on the SDSS spectroscopy. We have shifted the latter by about 0.1 dex
 to match the former. This conversion factor was obtained as in
Appendix A of \cite{Li2009} using a large sample of galaxies for which
both stellar mass estimates are available (see also Fig.~17
of \cite{Blanton2007}). The lensing and abundance matching results
agree well for stellar masses between $10^{10}M_{\odot}$ and
$10^{10.8}M_{\odot}$. At higher masses, the abundance matching halo
masses are somewhat larger than those estimated from lensing.

Triangles with error bars in Fig.~\ref{fig:mstar2mhaloobs} show mean
halo mass as a function of central stellar mass as derived from the
stacked kinematics of the satellites of a large sample of SDSS central
galaxies \citep{More2009}. Error bars here denote $1\sigma$
uncertainties in the mean. To make a direct comparison with our
predictions, we have converted the \cite{More2009} relation between
halo mass and galaxy luminosity into a relation between halo mass and
central stellar mass by using the SDSS data to estimate mean ``model''
stellar mass-to-Petrosian light ratio as a function of stellar
mass. We have also accounted for our differing definitions of ``halo
mass'' using an NFW profile of appropriate concentration. The
resulting $M_{halo}$ - $M_{*}$ relation is consistent with that found
from the lensing measurements, although its slope is slightly
flatter. Again, the $M_{halo}$ inferred at high mass is somewhat lower
than predicted by our abundance matching. 

We have studied whether this discrepancy could be due to a significant
dispersion in the central stellar mass of halos of given dark matter
mass. Assuming the dispersion in $\log M_*$ to be gaussian and
independent of $M_{halo}$, {\it rms} values exceeding about 0.2 are
excluded because they are inconsistent with the steep high-mass
fall-off of the stellar mass function of \cite{Li2009}.

In Fig.~\ref{fig:mstar2mhaloobs} we plot the mean halo masses
predicted for distributions which match both the SDSS/DR7 stellar mass
function and the halo mass function of Fig.~\ref{fig:MFwSat} assuming
dispersions in $\log M_*$ at given $M_{halo}$ of 0.1 (red curve) and
0.2 (green curve); our standard relation (the thick black curve)
corresponds to a dispersion of zero.  Dispersions in the allowed range
lead to a flatter slope of the $M_{halo}$ - $M_{*}$ relation and to
the prediction of a lower mean halo mass at high central stellar mass.
The effects are quite weak, however.

Fig. ~\ref{fig:mstar2mhaloobs} also shows the fraction of galaxies at
each stellar mass which are satellites according to our standard
assumption that $M_*$ is a monotonic function of $M_{halo}$ (the
dashed curve which is labelled on the right-hand axis). This fraction
maximises at about 40\% for the smallest galaxies considered in the
plot. Roughly 25\% of galaxies of Milky Way mass are satellites, but
less than 10\% of galaxies with $M_* > 10^{11.5}M_{\odot}$.  (These
fractions are consistent with those inferred using quite different
arguments by \cite{Mandelbaum2006}). These satellite fractions account
for at least partly the dispersion in the $M_{halo}$ - $M_{*}$
relation.

\subsection{Halo masses for galaxies in the Local Group }

\begin{table*}
\bc

\begin{tabular}{llllc}
  \hline
  {\bf Name} & $M_*$ & $M_{halo}$ & 80\% Confidence
  & Ref.\\
  & [$M_{\odot}$] & [$M_{\odot}$] &Interval [$M_{\odot}$] & \\
  \hline
  M31 &$6.98 \times 10^{10}$ &$2.96 \times 10^{12}$ &$[1.03,7.32] \times 10^{12}$   &(1) \\
  Milky Way & $5.5 \times 10^{10}$ &$1.99 \times 10^{12}$ &$[0.80, 4.74] \times 10^{12}$ & (2)\\
  M33 &$2.84 \times 10^{9}$ &$1.74 \times 10^{11}$&$[1.30, 2.23] \times 10^{11}$   &(1)\\
  LMC &$1.30 \times 10^{9}$ &$1.21 \times 10^{11}$  &$[0.93, 1.56] \times 10^{11}$ &(1)\\
  M32 &$1.24 \times 10^{9}$ &$1.19 \times 10^{11}$ &$[0.91, 1.53] \times 10^{11}$ &(1)\\
  NGC 205 & $9.29 \times 10^{8}$ &$1.05 \times 10^{11}$& $[0.81, 1.34] \times 10^{11}$ &(1)\\
  SMC &$2.63 \times 10^{8}$ &$6.39 \times 10^{10}$  &$[5.23, 7.91] \times 10^{10}$ &(1)\\
  &$5.95 \times 10^{8}$ &$8.69 \times 10^{10}$ &$[0.69,1.10] \times 10^{11}$ & (3)\\
  \hline
\end{tabular}

\caption{
  Stellar and halo masses for selected luminous Local Group galaxies.
  Stellar masses are computed from B-V colors and V-band magnitudes
  using the fit in table 1 of Bell \& de Jong for all galaxies except
  the Milky Way, for which $M_*$ is taken directly from Flynn et
  al. 2006.  The colors and magnitudes are taken from the references
  in column 5.  The halo masses in column 3 come from our abundance
  matching results (Fig. 2), while the values in column 4 give the
  10\% and 90\% values of $M_{halo}$ assuming $\sigma_{\log M_*}=0.2$.
  References: (1) de Vaucouleurs et al. 1991 (RC3); (2) Flynn et
  al. 2006; (3) Bothun \& Thompson 1988.  } \label{table:LGmasses}
\ec
\end{table*}

Within the Local Group a variety of dynamical tracers are available
which can provide halo mass estimates for individual galaxies.
Clearly, it is of interest to see how these compare with the halo
masses we infer from our abundance matching argument. A significant
obstacle to carrying through this programme is the difficulty of
estimating stellar masses for comparison to the more distant SDSS
galaxies. The large angular size of objects like M31, M33 and the
Magellanic Clouds makes it difficult to obtain accurate integrated
photometry, while our position within the Milky Way makes it
particularly difficult to infer our own Galaxy's total stellar mass.
In Table 1 we list stellar masses for the brighter galaxies of the
Local Group, together with the source from which we obtained them.
Where possible, we have chosen estimates based on similar techniques
and assumptions to those we use for more distant systems.

For given stellar mass, the $M_*$ -- $M_{halo}$ relation of
Fig.~\ref{fig:mstar2mhalo} predicts a unique value of $M_{halo}$.  We
list this for each galaxy in column 3 of Table 1. If this relation
does, in fact, have some dispersion, then a range of halo masses is
consistent with any given stellar mass. As noted above, the dispersion
in $\log M_*$ at given $M_{halo}$ cannot exceed 0.2 if the high-mass
tail of the SDSS/DR7 mass function is to be reproduced. Semi-analytic
simulations of galaxy formation within the
MS (e.g. DLB07) suggest a dispersion which is indeed
roughly independent of $M_{halo}$, but is somewhat smaller, $\sim
0.17$, at least for systems with $M_{halo} >10^{11}M_\odot$. At very
low masses, the models suggest a rather larger dispersion, $\sim 0.3$
for stellar masses around $10^8M_{\odot}$. \cite{More2009} infer a
very similar dispersion for the {\it luminosity} of central galaxies
in halos of given mass from their SDSS data on satellite
statistics. In column 4 of Table 1 we give the 10\% and 90\% points of
the distribution of halo mass predicted, given the stellar mass of
each galaxy, for a model with the maximal allowed dispersion (0.2) in
$\log M_*$ at given $M_{halo}$.

For the Milky Way, the recent stellar mass estimate
of \cite{Flynn2006} is quite similar to the old value of $\sim 6\times
10^{10}M_{\odot}$ found by 1980's models of Galactic structure. For
their estimated stellar mass of $5.5 \times 10^{10}M_{\odot}$, we
predict the halo mass of the Milky Way to be $2.0 \times
10^{12}M_{\odot}$. In a model with the maximally allowed dispersion,
the upper and lower 10\% points of the predicted halo mass
distribution are at $4.7 \times 10^{12}M_{\odot}$ and $0.80 \times
10^{12}M_{\odot}$. Most independent estimates of the halo mass of the
Milky Way have come from escape velocity arguments, or from Jeans
modelling of the radial density and velocity dispersion profiles of
distant halo tracer populations, e.g. red giants, blue horizontal
branch stars, globular clusters, or satellite galaxies. Recent studies
have typically come up with halo masses in the range $1$ to $2\times
10^{12} M_\odot$.  Estimates, in units of $10^{12}M_\odot$, include
$1.9^{+3.6}_{-1.7}$ \citep{Wilkinson1999}, $2.5^{+0.5}_{-1.0}$ or
$1.8^{+0.4}_{-0.7}$ \citep{Sakamoto2003}, depending on whether or not
Leo I is included \citep[see also,][ for the former case]{Li2008},
$1.42^{+1.14}_{-0.54}$ \citep{Smith2007},
$0.5-1.5$ \citep{Battaglia2005,Dehnen2006}, and
$1.0^{+0.3}_{-0.2}$ \citep{Xue2008}. These are consistent with
the values implied by our abundance-matching, though
some are at the lower end of the permitted range even when the $M_*$ --
$M_{halo}$ relation is allowed to have maximal scatter.

M31 appears to have a larger stellar mass than the Milky Way,
consistent with its larger maximum rotation velocity, and this
translates into a larger inferred halo mass $M_{halo} = 3.0\times
10^{12}M_{\odot}$. It is interesting that the sum of the halo masses
estimated for M31 and the Galaxy from our ``zero scatter'' abundance
matching is close to the best estimate of the same quantity
($M_{halo}(MW) + M_{halo}(M31) = 5.3\times 10^{12}M_{\odot}$)
which \cite{Li2008} obtained from a $\Lambda$CDM-calibrated Timing
Argument applied to the relative orbit of the two galaxies.  Note,
however, that if we allow maximal scatter in the $M_*$ -- $M_{halo}$
relation, then the Milky Way's halo mass could be as low as the values
found in other recent MW analyses, and the sum $M_{halo}(MW) +
M_{halo}(M31)$ would still not violate the 90\% confidence range
quoted by \cite{Li2008}.  In this case, the Milky Way's halo would of
course, be substantially less massive than those of typical galaxies
of similar stellar mass.

The other Local Group Galaxies listed in Table 1 are all predicted to
have (maximum past) halo masses at least a factor of 10 smaller than
those of the two giants.  As a result, they are likely to have caused
relatively little perturbation to the orbital dynamics of the main
binary system. The brightest of the satellites are nevertheless
predicted to have sufficiently massive halos that dynamical friction
may have modified their orbits. In addition, all of the galaxies show
evidence for tidal truncation (M32), tidal distortion (NGC205, M33),
or associated tidal streams (LMC, SMC, M33), so it is likely that
their current halo masses are smaller than the maximum values quoted
in Table 1.

\subsection{Stellar mass autocorrelations}

\begin{figure}
\resizebox{8.5cm}{!}{\includegraphics{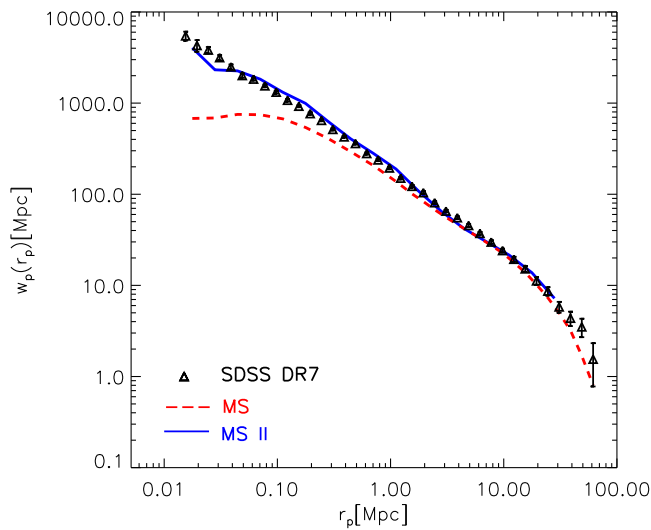}}\\%

\caption{The projected stellar mass autocorrelation function in the
SDSS/DR7 as measured by Li \& White (2009) is plotted as triangles with
error bars and is compared to the result obtained if $z=0$ (sub)halos
in the MS (dashed line) and the MS-II (solid line) are populated with
galaxies according to the $M_*$ - $M_{halo}$ relation of Fig. 2.}
\label{fig:wrp}
\end{figure}

In addition to estimating the stellar mass function of galaxies for
SDSS/DR7, \citet{Li2009} also studied the clustering of stellar mass
using the same galaxy sample.  This was quantified by the projected
autocorrelation function of stellar mass, $w^\ast_p(r_p)$. On scales
larger than individual galaxies $w_p^\ast(r_p)$ can be estimated with
high accuracy over about three orders of magnitude in $r_p$ and is
remarkably well described by a single power law. \cite{Li2009} showed
that this behaviour is approximately, but not perfectly reproduced by
existing galaxy formation simulations.

In Figure~\ref{fig:wrp} we show the predictions for $w_p^\ast(r_p)$
which result if (sub)halos in the MS and MS-II (dashed and solid
lines, respectively) are populated with galaxies according to the
$M_*$ -- $M_{halo}$ relation of Fig.~\ref{fig:mstar2mhalo}. We compare
these estimates to the SDSS/DR7 results of \citet{Li2009}.  Over the
range 20~kpc$<r_p<$20~Mpc, the MS-II prediction is in good agreement
with the SDSS data. The MS prediction converges to the MS-II on scales
larger than $\sim$2~Mpc, but is significantly too low on smaller
scales, becoming roughly constant for $r_p<100$~kpc. This reflects the
lower resolution of the MS. As noted above, it underpredicts (sub)halo
abundances for $M_{halo}<10^{12}M_\odot$ because many of these
correspond to satellite subhalos which have been stripped to masses
below the MS resolution limit. The objects missed are primarily in the
inner regions of massive halos, so their absence results in a
depression of small-scale clustering. In semi-analytic galaxy
formation simulations based on the MS, this effect is addressed by
explicitly following ``orphan'' galaxies from the time their subhalos
disrupt until the time that the code determines that they should
themselves disrupt or merge into the central galaxy \citep[see,
e.g.][]{Springel2005b, Croton2006}. This effect is negligible in MS-II
since, as noted above, the subhalo samples are essentially complete
down to $M_{halo}$ values that are small enough ($\sim
10^{10}M_{\odot}$) that their galaxies account for almost all
stars. Thus the excellent agreement between MS-II and the SDSS data
provides a powerful consistency check on the general framework
explored in this paper.

\subsection{Galaxy Formation Efficiency}
\begin{figure}
\bc
\hspace{-0.6cm}
\resizebox{8.5cm}{!}{\includegraphics{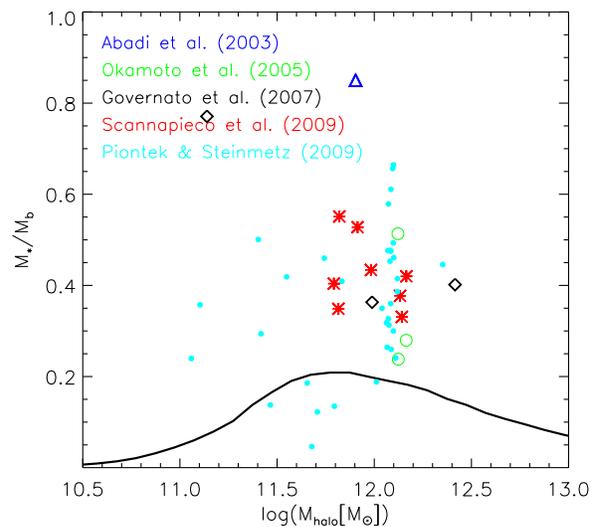}}\\%
\caption{Galaxy formation efficiency as a function of halo mass. The
black curve indicates the values required if a $\Lambda$CDM universe
is to fit the observed SDSS/DR7 stellar mass function. Coloured
symbols show the values found for a large number of recent simulations
of the formation of individual galaxies from $\Lambda$CDM initial
conditions.  Different colours correspond to simulations by different
authors as noted. The simulation results vary widely, but the great
majority lock too many baryons into stars to be viable models for the
bulk of the real galaxy population.}
\label{fig:mstar2mhalowobs}
\ec
\end{figure}

Given the relation between halo mass and stellar mass, we can define a
galaxy formation efficiency as the fraction of all baryons nominally
associated with the halo (calculated as the universal baryon fraction
times the halo mass) which are locked into stars. Thus,
\begin{equation}
{\rm Efficiency}
= \frac{M_*}{M_{halo}}\times \frac{\Omega_m}{\Omega_b} = 0.17 \times \frac{M_*}{M_{halo}}.
\end{equation}
We show this galaxy formation efficiency as a function of dark matter
halo mass in Fig.~\ref{fig:mstar2mhalowobs}. It peaks at around 20\%
in halos with $M_{halo}\sim 6\times 10^{11}M_{\odot}$, somewhat less
than the halo mass of the Milky Way. Similar numbers have previously
been derived from analogous arguments by \cite{Mandelbaum2006}
and \cite{Baldry2008}, among others. These low efficiencies must be
matched by galaxy formation simulations if these are to provide a
realistic description of the formation of real galaxies. In fact,
however, as shown by the coloured symbols in
Fig.~\ref{fig:mstar2mhalowobs}, most recent simulations of the formation
of galaxies of Milky Way mass convert 25\% -- 60\% of the available baryons into
stars \citep{Scannapieco2009, Governato2007,
Okamoto2005}. The efficiency in \cite{Abadi2003} is even higher, due to these
authors' neglect of SN feedback. Cyan dots show results of a survey of
baryonic physics parameter space by \cite{Piontek2009}. Several of their
models do show formation efficiencies as low as required to match the
SDSS stellar mass function in a $\Lambda$CDM universe, but the typical
value is around 35\%, almost twice as large as required.

Galaxy formation efficiency drops very rapidly towards both higher and
lower mass. In galaxy groups of mass $10^{13}M_{\odot}$, only 6\% of
the total baryons can condense to the centre and form stars.  This
reduction in efficiency may perhaps reflect the effects of feedback
from AGN \citep{Croton2006, Bower2006}. In halos of mass around
$4\times 10^{10}M_{\odot}$, around 1\% of the available baryons
have been converted into stars. Here, following the original
suggestion of \cite{Larson1974}, SN feedback is believed to be
responsible for the low efficiency, since it can expel gas effectively
from such shallow potential wells. In the smallest systems,
reionization may also play a role in suppressing condensation and star
formation \citep{Efstathiou1992, Benson2002, Sawala2009}.

\subsection{The stellar mass ``Tully-Fisher'' relation}

\begin{figure}
\bc
\hspace{-0.6cm}
\resizebox{8.5cm}{!}{\includegraphics{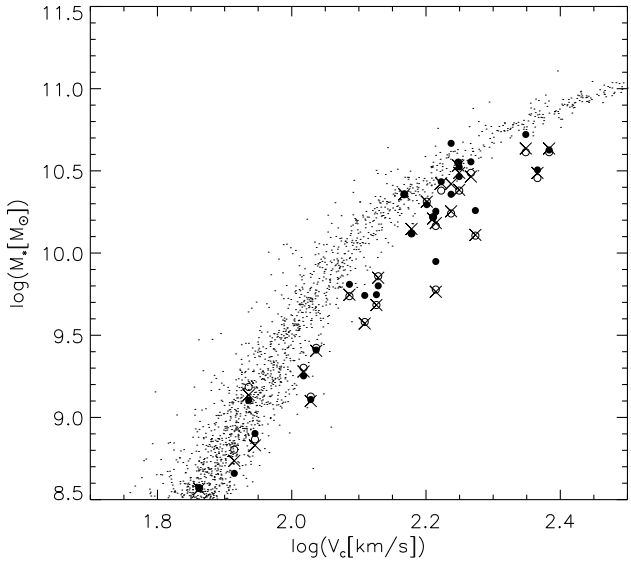}}\\%
\caption{The stellar mass ``Tully-Fisher'' relation. Small black dots
plot the stellar masses predicted for individual MS-II halos by the
$M_*$ -- $M_{halo}$ relation of Fig. 2 against their current maximum circular
velocity. Big symbols are based on estimates of the stellar mass of
real galaxies from I band (crosses), K band (filled circles), and B
and R band (open circles) photometry and on direct measures of their
maximum rotation velocity \citep{Bell2001}.}
\label{fig:tf}
\ec
\end{figure}

A long standing problem in $\Lambda$CDM cosmology has been to
reproduce simultaneously the galaxy luminosity function and the
zero-point of the Tully-Fisher
relation \citep{Kauffmann1993,Cole1994,Navarro2000,Cole2000}. The
abundance matching method used here reproduces the observed stellar
mass function automatically. To establish a link between stellar mass
and circular velocity, we use the $M_*$ -- $M_{halo}$ relation to
assign a stellar mass to the central galaxy of each dark matter
(sub)halo in the MS-II. The maximum circular velocity of these
subhalos is tabulated in the simulation database, and we take this as
a proxy for the maximum rotation velocity of the galaxy. The resulting
``Tully-Fisher'' relation is shown in Fig.~\ref{fig:tf}. Black dots
are our predictions for central galaxies.  There is a tight relation
between stellar mass and halo maximum circular velocity which can be
described approximately by a double power law. The bend corresponds to
the turn-over point in the $M_*$ -- $M_{halo}$ relation
(Fig.~\ref{fig:mstar2mhalo}), and to the point where galaxy formation
efficiency reaches its maximum. Large symbols show data for observed
spiral galaxies taken from \cite{Bell2001} who adopted a scaled-down
Salpeter Initial Mass Function (IMF) when deriving their stellar
masses. This gives values which are higher by 0.15 dex than those
used here, which assume a Chabrier IMF. We shift the observational
data downwards by 0.15 dex in order to compensate for this difference.

The $\Lambda$CDM model predicts circular velocities which are similar
to or lower than those observed over the full stellar mass
range. Moderate differences in this sense are expected, since the
simulations do not account for the gravity of the baryons (see the
discussion in \cite{Navarro2000}). In the region $2.0< \log V_{c}<2.2$
where spiral galaxies dominate the mass functions, the predicted
circular velocity at each stellar mass is lower than in the
observations by about 25\%. This is plausible, given results from
detailed simulations of spiral
formation \citep[e.g.][]{Gustafsson2006,Abadi2009}. These simulations show that galaxy condensation leads to a compression of the inner dark halo which is similar to but somewhat weaker than that predicted by simple adiabatic contraction models \citep{Barnes1984, Blumenthal1986, Gnedin2004}. The combined effect of the baryonic galaxy and the compressed dark halo is an increase in maximum circular
velocity which may be of the order we require. Such an enhancement is
already included approximately in many disk galaxy formation
models \citep[e.g.][]{Mo1998, Cole2000}.  Recent
studies \citep{Gnedin2007, Dutton2007} suggest that still larger
circular velocity enhancements may be produced, in which case the
galaxy luminosity function and Tully-Fisher relation cannot be matched
simultaneously, but this depends on the details of galaxy formation
and assembly and goes beyond the issues we can discuss here. At higher masses $\log V_{c}>2.2$, the difference between prediction and
observation is smaller, though the model still predicts a slightly
lower circular velocity for a given stellar mass. This again
corresponds well to observation since strong gravitational lensing
studies suggest that the circular velocity is constant at close to the
halo maximum value throughout the central regions of these higher mass
galaxy-halo systems \citep{Gavazzi2007}. At the low-mass end, $\log
V_{c}<2.0$, dark matter dominates the gravity throughout observed
galaxies and the model prediction matches observation rather well. In
summary, at given stellar mass, circular velocities are expected to be
higher than represented by the small dots in Fig.~\ref{fig:tf} because
of the gravitational effects of the baryons. The observed Tully-Fisher
relation could nevertheless be reproduced if the increase in circular
velocity is typically $\sim$20\%, a value which is not implausible
given current simulation data.

\section{Conclusions}
We have related the stellar mass of a galaxy to the dark matter mass
of its halo by adopting a $\Lambda$CDM cosmology and assuming that the
stellar mass of a galaxy is a scatter-free and monotonic function of
the maximum mass ever attained by its halo. By combining the MS and
the MS-II we are able to derive accurate halo/subhalo abundances for
maximum masses spanning the entire relevant range, $10^{10}M_{\odot}$
up to $10^{15}M_{\odot}$. The SDSS/DR7 provides precise galaxy
abundances spanning almost 4 orders of magnitude in stellar mass. By
comparing the two we have linked stellar mass to halo mass with high
formal accuracy over this full stellar mass range without any further
assumptions about galaxy formation physics or halo evolution.

The ratio of stellar mass to halo mass maximises at about 3\% for
galaxies somewhat fainter than $L^*$, and decreases rapidly towards
both higher and lower mass. Except possibly for the most massive
galaxies, halo masses derived from this abundance matching argument
agree well with those estimated from weak gravitational lensing of
background galaxies and from a stacking analysis of SDSS satellite
dynamics. We investigated whether the discrepancy at high mass might
reflect a dispersion in the stellar mass of central galaxies at fixed
halo mass, and found that the maximal allowed dispersion of 0.2 in
$\log M_*$ at given $M_{halo}$ leads to somewhat better agreement
between our prediction and observations.

Using our $M_*$ -- $M_{halo}$ relation we have predicted halo masses
for a number of the more massive Local Group galaxies.  For the Milky
Way the inferred halo mass is around $2\times10^{12}M_{\odot}$,
consistent with most recent estimates from the dynamics of halo
tracer populations.  The inferred halo mass for M31 is larger, around
$3 \times 10^{12}M_{\odot}$. The sum of the two is in excellent
agreement with the value found when the Timing Argument is applied to
the relative motion of the two giants. The halo masses of the brighter
Local Group satellites are all inferred to be less than 10\% of those
of the giants, and so should have only rather minor effects on Local
Group dynamics.

Galaxy formation efficiency peaks at $\sim$ 20\% in halos slightly
less massive than the hosts of $\sim L^*$ galaxies. It drops rapidly
at both higher and lower mass. Similar values have been derived
previously by others from weak lensing and abundance matching
studies. Comparison with recent hydrodynamic simulations of galaxy
formation shows that most simulations have conversion efficiencies
which are too high for them to be viable models for the bulk of the
real galaxy population.

The stellar mass -- halo mass relation which we derived from our
abundance matching argument can be used to populate halos in the
Millennium and Millennium-II simulations. For the latter, the implied
spatial clustering of stellar mass is in remarkably good agreement
with a direct and precise measurement based on the full SDSS/DR7
dataset. Clustering is underpredicted on scales below a Mpc in the
Millennium Simulation because subhalos corresponding to relatively
massive satellite galaxies on tightly bound orbits are often missed at
MS resolution. This effect is mitigated in galaxy formation
simulations based on the MS by following ``orphan'' galaxies after
their subhalos have disrupted.

The maximum circular velocity of each subhalo in the MS-II can be
identified with the maximum rotation velocity of the central galaxy
assigned to it by the above procedure. This results in a stellar mass
``Tully-Fisher'' relation which we studied over the rotation velocity
range $1.8<$log$(V_c[km/s])<2.5$. For galaxies like the Milky Way,
this model predicts circular velocities which are about 20\% lower
than observed. This is roughly consistent with the difference expected
due to the neglect of the gravitational effects of the stars. At lower
masses dark matter dominates throughout the galaxies, and our results
match the observations quite well. Thus, the $\Lambda$CDM cosmology
does seem able to reproduce observed luminosity functions and
Tully-Fisher relations simultaneously.

Although the abundance matching scheme is a powerful way to relate
galaxies to their dark matter halos, in reality, there must be some
scatter in the relation which is likely to depend on other physical
properties of the systems. An important source of scatter is the
evolution of the stellar mass -- halo mass relation. The gas fraction
of galaxies is very likely greater at high redshift than in the local
universe. More galactic baryons may be in the form of gas than in
stars during the early stages of galaxy
formation. Fig.~\ref{fig:mstar2mhaloobs} shows that the satellite
fraction is higher for smaller stellar mass, suggesting there may be
more scatter in low-mass galaxies than at high mass. Additional
scatter may come from variations in halo assembly
history \citep[e.g.][]{Croton2007,Gao2007}. For example, the star
formation efficiency is higher in merger-induced bursts than in
quiescent phases. We can study such effects by comparing observations
to models for galaxy correlations as a function of redshift and of
stellar mass and age. Direct and precise measurement of the stellar
mass function at high redshift will also help us study the scatter
more quantitatively, because the way in which galaxies populate halos
at different redshifts is tightly coupled to the evolution of the dark
halo distribution, and so to the merger trees we construct for our
simulations.
 
\label{sec:conclusiongf}

\section*{Acknowledgements}
 We thank Maiangela Bernardi and Francesco Shankar for useful
 discussion of stellar mass estimates. The MS and MS-II simulations
 used in this paper were carried out at the Computing Centre of the
 Max Planck Society in Garching. The halo data are publicly available
 at http://www.mpa-garching.mpg.de/millennium/.

\section*{Appendix A}
The stellar mass estimates in \cite{Li2009} were based entirely on the
SDSS Petrosian magnitudes. For each galaxy these are measured for all
five bands within the same circular aperture, defined to have radius
twice the galaxy's measured Petrosian radius. This has the advantages
that the colours then refer to a specific and well defined stellar
population, and that no extrapolation from the direct measurements is
involved. On the other hand, some light falls outside this aperture so
that the Petrosian magnitudes underestimate the total luminosities of
galaxies by amounts which depend on the shapes of their individual
surface brightness profiles.  For example, the SDSS Petrosian
magnitudes recover almost all the flux from objects with exponential
profiles, but only about 80\% of the flux from objects with de
Vaucouleurs profiles. In the procedure of \cite{Blanton2007} the
Petrosian colours of each galaxy are fit to stellar population
synthesis templates shifted to its known redshift. For an assumed
initial mass function (here that of \cite{Chabrier2003}) this gives a
stellar mass-to-light ratio for the galaxy.  Multiplying by the
Petrosian luminosity then gives the stellar mass estimate.

In this paper we are concerned to describe the relation between the
total stellar mass of galaxies and their halo mass. In addition to
Petrosian magnitudes, the SDSS databases give ``model'' magnitudes for
each galaxy. Exponential and de Vaucouleurs models are convolved with
the point-spread-function and fit to the $r$-band aperture
photometry. The total magnitude corresponding to the best fit of the
two is then defined as the $r$-band model magnitude of the object.
This clearly involves extrapolation from the measurements, and in
individual cases the errors can be significant. Nevertheless, numerous
tests suggest that in general this produces unbiased and reasonably
robust estimates of total luminosity\footnote{See the discussion at\\
http://www.sdss.org/dr7/algorithms/photometry.html}. We use these
model magnitudes to correct the ``Petrosian masses'' used
by \cite{Li2009} to ``total masses'' by multiplying the former by the
ration of the fluxes corresponding to the $r$-band model and Petrosian
magnitudes. Note that this procedure ensures that the colours fit to
the population sysntehsis templates do indeed refer to a well defined
stellar population which encompasses the bulk of each galaxy's stars.
This would not be the case if we had instead used ``model'' colours
directly.

In fig.~A1 we compare the ``total'' stellar mass function (red) obtained in
this way to the ``Petrosian'' stellar mass function (black) of \cite{Li2009}.
We only plot masses above $10^{10}M_\odot$ since the two functions are
indistinguishable at lower mass. It is interesting that the
differences are largest at the highest masses, reflecting the fact
that massive galaxies tend to have de Vaucouleurs rather than
exponential profiles.  Indeed, the highest mass systems are often cD's
with extended envelopes which rise above a de Vaucouleurs fit to their
high surface brightness regions. The stellar masses of such systems
will still be systematically underestimated by our ``total'' masses.

As shown in Fig.A1, except at the highest masses, the difference
between the two mass functions is quite well represented by a shift in
mass of 0.04 dex.  A slightly better representation of the ``total''
stellar mass function is obtained by modifying the model given
in \cite{Li2009} to have the parameters listed in Table 2.

\begin{table*}
\bc
\caption{Parameters of a triple Schechter function fit to the ``total'' stellar mass function.}

\begin{tabular}{cccc}
  \hline        
  $\mathrm{Mass}$ $\mathrm{range}$  & $\Phi^*$ & $\alpha $ & $\log _{10}M^*$ \\
  $\mathrm{[}M_{\odot}\mathrm{]}$ & [$h^3 \mathrm{Mpc}^{-3}\log _{10} M_{\odot}^{-1}$] & &[$M_{\odot}$]  \\
  \hline
  $8.27 < \log _{10} M < 9.60$ &$0.159(5)$ &$ -1.11(09)$ &$9.84(21)$    \\
  $9.60 < \log _{10} M < 10.94$ & $0.0121(7)$ & $-0.938(035)$ &$10.71(03)$ \\
  $10.94 < \log _{10} M < 12.27$ &$0.0032(5)$ & $-2.33(11)$&$11.09(03)$   \\  
  \hline
\end{tabular}
\ec
\end{table*}

\begin{figure}
\bc
\hspace{-0.6cm}
\resizebox{8.5cm}{!}{\includegraphics{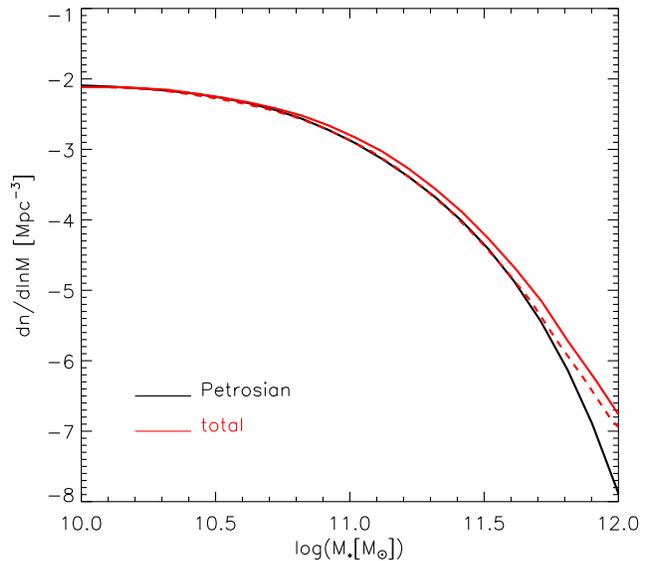}}\\%
\caption{A comparison of the stellar mass function of Li \& White (2009),
which was based on the SDSS $r$-band Petrosian magnitudes, to a
similarly calculated ``total'' stellar mass function based on the SDSS
$r$-band model magnitudes. The dashed line shows the latter function
shifted to smaller masses by 0.04 dex.  Except at the highest masses,
this is a good representation of the difference between the two mass
functions.}
\label{fig:MFs}
\ec
\end{figure}

\bibliographystyle{mn2e}

\bibliography{g2h}

\end {document}